\documentstyle[prb,aps,twocolumn,floats]{revtex}
\newcommand{\be}{\begin{equation}}
\newcommand{\ee}{\end{equation}}
\input psfig.sty
\begin{document}
\draft
\title{Single-hole dynamics in dimerized and frustrated spin-chains}
\author{Christoph Jurecka and Wolfram Brenig}

\address{Institut f\"ur Theoretische Physik,
Technische Universit\"at Braunschweig, 38106 Braunschweig, Germany}

\date{\today}
\maketitle
\begin{abstract}
We present a unified account for the coupled single-hole- and spin-dynamics
in the spin-gap phase of dimerized and frustrated spin-chains and two-leg
spin ladders. Based on the strong dimer-limit of a one-dimensional
$t_{1,2,3}$-$J_{1,2,3}$-model a diagrammatic approach is presented which
employs a mapping of the spin-Hamiltonian onto a pseudo-fermion bond-boson
model. Results for the single-hole spectrum are detailed. A finite
quasi-particle weight is observed and studied for  a variety of system
parameters. A comparison with existing exact diagonalization data is
performed and good agreement is found.
\end{abstract}


\section{Introduction}\label{sec1}
Unconventional quantum-magnetism in low-di\-men\-si\-onal transition-metal
compounds has received considerable interest recently due to the discovery
of numerous novel materials with spin-$\frac{1}{2}$ moments arranged in
chain, ladder, and depleted planar structures.  Among these compounds are
antiferromagnetic chain-systems which are intrinsically dimerized, in
particular, (VO)$_2$P$_2$O$_7$ \cite{Garre97a}, CuWO$_4$ \cite{Lake96}, and
Cu(NO$_3$)$_2\cdot2.5$H$_2$O \cite{Bonne83}. Moreover, 
quasi-one-dimensional (1D) materials have been discovered which display 
a temperature dependent dimerization, eg. CuGeO$_3$ \cite{Hase93a} 
and $\alpha'$-NaV$_2$O$_5$
\cite{Isobe96} where the former is the first inorganic spin-Peierls chain
\cite{Bulaevskii78} and magnetically frustrated \cite{Castilla95,Riera95}
while the latter is a $\frac{1}{4}$-filled two-leg ladder
\cite{Smolinski98}.  Spin-ladders, both inorganic, eg. SrCu$_2$O$_3$
\cite{Azuma9194} and CaV$_2$O$_5$ \cite{Iwase96} as well as organic, eg.
Cu$_2$(C$_2$H$_{12}$N$_2$)$_2$Cl$_4$ \cite{Chabo97} have been investigated.
Quite recently SrCu$_2$(BO$_3$)$_2$ \cite{Kageyama99,Miyahara99} has been
shown to realize the two-dimensional (2D) version \cite{Shastry81} 
of the 1D frustrated Majumdar-Ghosh model \cite{Majumdar69a}.
 
In contrast to conventional 1D spin-chain materials, eg.  Sr$_2$CuO$_3$ and
SrCuO$_2$ \cite{Motoyama96,Teske6970}, which display algebraic, almost
long-range, spin-correlations and gap-less magnetic excitations the new
materials are spin-liquids (or dimer-solids) with short-range singlet
correlations and a gap in the spin spectrum. The spin-gap phenomenon has
been attributed to dimerization \cite{Bulaevskii78,Cross79a} and 
frustration \cite{Majumdar69a,Okamoto92a,Chitra95a} in chain-systems,
 and it can be interpreted accordingly in ladders due to their topological
equivalence to frustrated and dimerized chains \cite{LadderRev,ElbioRev}.

Apart from magnetic excitations the dynamics of electronic carriers doped
into quasi-1D spin-liquids is of interest, in particular because of the
discovery of superconductivity \cite{Uehara96} in the two-leg
ladder-compound Sr$_{14-x}$Ca$_x$Cu$_{24}$O$_{41}$ \cite{Carron88}.
Important progress has been achieved regarding the pairing correlations and
the phase diagram of spin-ladders \cite{ElbioRev}. However, only a
restricted set of primarily numerical studies
\cite{Tsunet94,Tsunet95,Troyer96,Haas95,Haas96,augier97,eder98,Martins99} 
has focused on the spectral properties of single-hole excitations in
quasi-1D spin-liquids at low doping. In this regime, and due to the
spin-gap, a quasi-particle picture \cite{Luther74}, rather than a
Luttinger-liquid description \cite{Haldane80}, is believed to be 
relevant at low energy scales \cite{Tsunet95}.

In this paper we detail a theory of single-hole excitations at half-filling
for a dimerized and frustrated spin-chain. Particular emphasis will be on
the coupling between spin and charge degrees of freedom. The paper is
organized as follows: first we introduce the $t_{1,2,3}$-$J_{1,2,3}$-model
for the dimerized and frustrated spin system. Second we summarize a
bond-boson-method to evaluate the spin-excitations at half filling starting
from the strong dimer limit. Third, we map the 
$t_{1,2,3}$-$J_{1,2,3}$-model
in the single-hole sector onto a coupled boson-fermion-model. Next, the
single-hole excitations of the latter model are evaluated by a
selfconsistent diagrammatic technique.  Finally we present results for the
spectral properties of the single hole, both for ladders and chains, and
compare our findings to existing numerical analysis of finite systems.

\section{The $\lowercase{t}_{1,2,3}$-$J_{1,2,3}$-model}
\begin{figure}[th]
\psfig{file=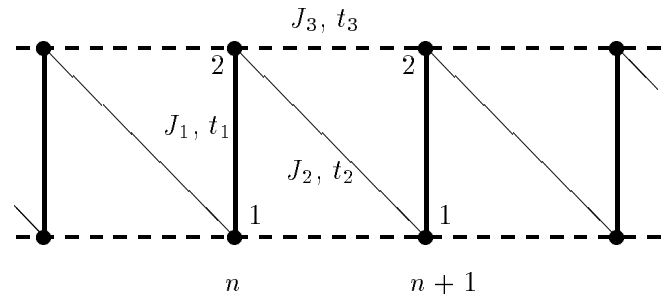,width=\columnwidth}
\vspace{0.3cm}
\caption[l]{$t_{1,2,3}$-$J_{1,2,3}$-model. $n$ labels the dimer
bonds. 1 and 2 refer to non-equivalent sites per unit cell.}
\label{fig1}
\end{figure}

For the remainder of this work we focus on systems which are of a Mott- or
charge-transfer type of insulator at half filling and allow for an
approximate mapping to a $t$-$J$-model. To describe the case of either
dimerization and frustration of the magnetic exchange coupling along the
chain or the topology of a two-leg ladder we invoke a $t$-$J$-model which
includes hopping- and exchange-integrals up to third-nearest-neighbors
(NNN), i.e.  $t_{1,2,3}$ and $J_{1,2,3}$, as shown in figure \ref{fig1}.

Depending on the case of a ladder or a chain the real-space structure 
of the system is either identical to that of figure \ref{fig1} 
with $J_2=t_2=0$ or
can be obtained by deforming the figure such as to place sites $2,n$
half-way in-between sites $1,n$ and $1,n+1$. The Hamiltonian reads
\begin{eqnarray}\label{e1}
H=&&H_t+H_J
\\\nonumber\\ 
H_J=&&J_1 \sum_n {\bf S}_{1, n} {\bf S}_{2, n}+J_2 \sum_n {\bf S}_{2, n}
{\bf S}_{1, n+1}
\nonumber\\
&&+ J_3 \sum_i ({\bf S}_{1,n} {\bf S}_{1,n+1}+{\bf S}_{2,n} 
{\bf S}_{2, n+1}),
\label{e2}\\
H_t=&&-t_1\sum_{n, \sigma} \hat{c}_{1, n, \sigma}^\dagger \hat{c}_{2, n, 
\sigma}^{\phantom\dagger}-t_2\sum_{n, \sigma}
 \hat{c}_{2, n, \sigma}^\dagger 
\hat{c}_{1, n+1, \sigma}^{\phantom\dagger}
\nonumber\\
&& -t_3\sum_{n, \sigma}(\hat{c}_{1, n
\sigma}^\dagger \hat{c}_{1, n+1, \sigma}^{\phantom\dagger}+\hat{c}_{2, n
\sigma}^\dagger \hat{c}_{2, n+1, \sigma}^{\phantom\dagger})+h.c.
\label{e2a2}
\end{eqnarray}
where ${\bf S}_{i,n}$ are spin-1/2 operators on site $i$ of dimer-bond $n$
and $\hat{c}_{i,n, \sigma}^{(\dagger)}=[c_{i,n, \sigma} (1-n_{i,n,
-\sigma})]^{(\dagger)}$ are projected fermion-operators of spin $\sigma$ on
site $i,n$. Dimerization and frustration of the 
spin-system is expressed via
the parameters $\delta$ and $\alpha$ where $J_{1(2)}=J_0 (1+(-)\delta)$ and
$\alpha=J_3/J_0$. Similarly the hopping integrals are related through
$t_{1(2)}= t_0 (1+(-)\tilde{\delta})$ and $\tilde{\alpha}=t_3/t_0$. Ladders
are characterized by $\delta=\tilde{\delta}=1$. Regarding the spin-part,
i.e. $H_J$, the ground state is known to be a product-state of singlets
located on the dimer-bonds for $\alpha$ and $\delta$ on the disorder-line
\cite{Majumdar69a,Shastry81} $\delta=1-2\alpha$, moreover, 
the spin-spectrum
is gap-less only at $\delta=0$ and $0\leq\alpha<\alpha_C$ with
$\alpha_C\simeq 0.2411$ \cite{Okamoto92a,Chitra95a}. For magnetic couplings
which are mediated by super-exchange and for systems with only a single
relevant on-site Coulomb-energy scale we expect that $(1+\tilde{\delta})^2
\sim(1+\delta)$ and $\tilde{\alpha}^2\sim\alpha$ 
for $|\delta,\alpha|\ll 1$.

\section{Undoped spin-system}
The spin-part, i.e. $H_J$, of Hamiltonian (\ref{e1}) allows for a mapping
onto a model of hard-core 'bond'-bosons originally employed in the context
of the 2D Heisenberg-model for the high-T$_C$ cuprates \cite{Sachdev90a}.
Here we briefly restate the essential features of this mapping. 
Consider any
two spin--1/2 operators ${\bf S}_{1,n}$ and ${\bf S}_{2,n}$. 
The eigenstates
of the related total spin are a singlet and three triplets These can be
created out of a vacuum $\left|0\right>$ by applying the bosonic operators
$s_{n}^\dagger$ and $t_{\alpha,n}^\dagger$ with $\alpha= x,y,z$
\begin{eqnarray}
s_n^\dagger|0\rangle = \frac{1}{\sqrt{2}}\left(|\!\uparrow
\downarrow\rangle-|\!\downarrow\uparrow\rangle\right)_n\nonumber \\
t_{x,n}^\dagger|0\rangle = \frac{-1}{\sqrt{2}}\left(|\!\uparrow
\uparrow\rangle-|\!\downarrow\downarrow\rangle\right)_n\nonumber \\
t_{y,n}^\dagger|0\rangle = \frac{i}{\sqrt{2}}\left(|\!\uparrow
\uparrow\rangle+|\!\downarrow\downarrow\rangle\right)_n\nonumber \\
t_{z,n}^\dagger|0\rangle = \frac{1}{\sqrt{2}}\left(|\!\uparrow
\downarrow\rangle+|\!\downarrow\uparrow\rangle\right)_n
\label{e4}
\end{eqnarray}
where the first (second) entry in the kets refers to site 1(2) on dimer $n$
of fig. \ref{fig1}. On each site we have $[s,s_{\phantom{\alpha}
}^\dagger]=1$, $[s_{\phantom{\alpha}}^{(\dagger)},t_\alpha^{(\dagger)}]=0$,
and $[t_\alpha^{\phantom{(\dagger)}},t_\beta^{\dagger}]=\delta_{\alpha
\beta}$. The action of ${\bf S}_{1,n}$ and ${\bf S}_{2,n}$ in this
space leads to the representation
\be
S^\alpha_{^1_2,n}=\frac{1}{2}(\pm 
s_n^\dagger t_{\alpha, n}^{\phantom\dagger} \pm t_{\alpha, n}^\dagger 
s_n-i\epsilon_{\alpha\beta
\gamma}^{\phantom\dagger} t^\dagger_{\beta, n}
t_{\gamma,n}^{\phantom\dagger}).
\label{e3}
\ee
Here $\varepsilon_{\alpha\beta\gamma}$ is the Levi--Civita symbol and a
summation over repeated indices is implied hereafter.  
The upper(lower)
subscript on the lhs. of (\ref{e3}) refer to upper(lower) sign on the rhs..
The bosonic Hilbert space has to be restricted to the physical Hilbert
space, i.e. to either one singlet or one triplet per site, by the 
constraint
\be\label{b3}
s_n^\dagger s_n^{\phantom\dagger}+t_{\alpha,n}^\dagger t_{
\alpha,n}^{\phantom\dagger}=1
\label{e5}
\ee
The representation (\ref{e3}) can be inserted into (\ref{e2}) yielding an
interacting bose-gas Hamiltonian accompanied by the constraint (\ref{b3})
\cite{Sachdev90a}. This Hamiltonian is diagonal intra-dimer-wise and
contains two-particle interactions which are only of inter-dimer type.  At
the point of complete dimerization, i.e., $\delta=1$ and $\alpha=0$, the
inter-dimer interactions vanish leaving a sum of purely local dimer
Hamiltonians, each of which has a singlet ground state. This renders the
global ground-state a product of singlets localized on the dimers with the
excitations being a set of 3$^N$-fold degenerate triplets. Off the
dimer-point the inter-dimer interactions can be treated approximately by a
linearized Holstein-Primakoff (LHP) approach, details of which can be found
in the literature \cite{Chubokov89a,Chubukov91a,Starykh96a,brenig98}. The
LHP method retains spin-rotational invariance and reduces $H_J$ to a set of
three degenerate massive magnons
\be
H_J=\sum_k \omega_k \gamma_{\alpha,k}^\dagger
\gamma_{\alpha,k}^{\phantom\dagger} + \mbox{const.}
\label{e7}
\ee
with
\begin{eqnarray}
t_{\alpha,k}^\dagger & = & u_k \gamma_{\alpha,k}^\dagger+v_k \gamma_{
\alpha,-k}^{\phantom\dagger},
\label{e6}\\
\omega_k  & = & J_1 \sqrt{1+2 e_k}\label{e8}
\label{e66}\\
e_k & = &\frac{2 J_3-J_2}{2 J_1} \cos{k}
\label{e9}\\
u[v]_k^2
& = & \frac{1}{2}\left(\frac{J_1 (1+ e_k)}{\omega_k}+[-]1\right)
\label{e10}
\end{eqnarray}
The '$[]$'-bracketed sign on the rhs. in (\ref{e10}) refers to the quantity
$v$ on the lhs.. The spin-gap $\Delta=\min\{\omega_k\}$ resides at
$k=\pi(0)$ with $\Delta=\sqrt{J_1^2-(+)J_1(2J_3-J_2)}$ for $2J_3>(<)J_2$.
Note that because of (\ref{e6}) the ground state $|D\rangle$, which is
defined by $\gamma_{\alpha,k}|D\rangle=0$, contains quantum-fluctuations
beyond a pure singlet product-state.

To leading order the dispersion $\omega_k$ is identical to perturbative
expansions, both, for chains \cite{Harris73a,Uhrig97a} and ladders
\cite{Reigrotzki94,Barnes93} and it has been applied to model inelastic
neutron scattering (INS) data for CuGeO$_3$ \cite{brenig98} and
$\alpha'$-NaV$_2$O$_5$ \cite{Gros99}.  Beyond the LHP approach
triplet-interactions lead to a renormalization of $\omega_k$ and to the
formation of multi-magnon bound states \cite{Kotov98,Kotov99,Jurecka00}.
Moreover the constraint (\ref{e5}), although a hardcore repulsion, has been
considered perturbatively relying on the case of low triplet-density at
$\Delta/J\ll 1$ \cite{Kotov98,Kotov99,Kotov98-2}. These renormalizations of
the triplet dispersion will be discarded in our evaluation of the
single-hole spectra, in particular since their dominant effects can be
accounted for semi-phenomenologically by adjusting the size of the LHP
spin-gap. This is of significance when comparing single-hole spectra to
numerical studies, as in section \ref{sec7}.

\section{Single-hole Hamiltonian}
In this section we map the $t_{1,2,3}$-$J_{1,2,3}$-Hamiltonian of
(\ref{e1}-\ref{e2a2}) in the single-hole sector onto a model of
pseudo-fermions interacting with the bond-bosons. To this end we note 
that a dimer-bond occupied by a single hole can be labeled by 
introducing an additional pseudo-fermion (holon).  I.e., instead of 
the bond being in one of the states given by (\ref{e4}) it can also 
be in the state
\begin{eqnarray}\label{e11}
a_{j,n,\sigma}^\dagger|0\rangle=|j\sigma\rangle_n
\end{eqnarray}
where the l.h.s. denotes the vacuum, i.e. $|0\rangle$, with a single
dimer-bond at site $n$ in a one-hole state of spin $\sigma$ with $j=1,2$
referring to the two positions available to the hole on the bond. The
operators  $a^{\phantom{\dagger}}_{j, i, \sigma}$ are required to obey
fermionic anticommutation relations. In the single-hole sector each
dimer-bond can only be in exactly one of the states given by (\ref{e4}) or
(\ref{e11}). Therefore an extended hard-core constraint has to be satisfied
\begin{equation}
s_i^\dagger s_i^{\phantom\dagger}+t_{\alpha, i}^\dagger 
t_{\alpha, i}^{\phantom\dagger}+
\sum_{j=1,2} a_{j,i, \sigma}^\dagger a_{j,i, \sigma}^{\phantom\dagger}=1
\label{e18}
\end{equation}
with a summation over repeated spin-indices implied.  Creation of a single
{\em physical hole} in the (half-filled) ground state $|D\rangle$ of the
spin-system is achieved by applying the {\em two-particle} operator
\begin{eqnarray}\label{e13}
\hat{c}_{j, n, \sigma} =&& \frac{p_j}{\sqrt{2}}[
a_{\overline{j}, n, \overline{\sigma}}^\dagger
(p_j p_\sigma s_n^{\phantom\dagger} + t_{z,n}^{\phantom\dagger})
\nonumber\\
&& +a_{\overline{j}, n, \sigma}^\dagger
(p_{\overline{\sigma}} t_{x,n}^{\phantom\dagger}
+i t_{y,n})^{\phantom\dagger}]
\end{eqnarray}
where $p_j=+(-)$, $\overline{j}=2(1)$ for $j=1(2)$ and $p_\sigma= +(-)$,
$\overline{\sigma}=\downarrow(\uparrow)$ for 
$\sigma=\uparrow (\downarrow)$.
In terms of the previous equation the creation of a physical hole can be
interpreted has the removal of a spin-dimer followed by the creation of
dimer hole-state, where the particular linear combination of the $a_{j,
n,\sigma}^\dagger$, $s_n^{\phantom\dagger}$, and
$t_{\alpha,n}^{\phantom\dagger}$ operators on the r.h.s. ensures that the
total spin is $S=1/2$ and $S_z=\pm 1/2$. Using the constraint (\ref{e18}) 
it is straightforward to show, that the r.h.s. of (\ref{e13})
indeed satisfies the usual Hubbard-operator algebra.

Inserting (\ref{e13}) into (\ref{e1}-\ref{e2a2}) we have to distinguish
between various cases, i.e., (i) intra-dimer hopping, (ii) inter-dimer
hopping, and (iii) exchange-scattering. Inter-dimer hopping can be either
spin-diagonal or accompanied by spin-flip scattering. This includes
(ii,a) singlet-singlet, (ii,b) singlet-triplet, and triplet-triplet
transitions of the spin-background upon hole-hopping, the latter may occur
with a change in the spin-quantum number of the background of either
(ii,c): $\Delta S_z=0$ or (ii,d): $\Delta S_z= 1$.

Processes of the type (i) and (ii,a-d) result from direct insertion of
(\ref{e13}) into (\ref{e2a2}). To express the inter-dimer
exchange-scattering, i.e. process (iii), one has to realize that the spin
operator on a dimer in a one-hole state needs to be expressed in terms of
the pseudo-fermions rather than the bond-bosons, i.e. using
\be
{\bf S}_{m,n}^{ij}=\frac{1}{2}
\sum_{\sigma_1, \sigma_2} a_{i,m,\sigma_1}^\dagger
 {\boldmath\tau}_{\sigma_1
\sigma_2} a_{j,n,\sigma_2}^{\phantom\dagger}
\label{e17}
\ee
and setting $j=i$ and $m=n$, where ${\boldmath\tau}_{\sigma_1 \sigma_2}$
are the Pauli matrices and $\left ({\bf S}_{m,n}^{ij}\right)^\dagger= {\bf
S}_{n,m}^{ji}$.  Substituting this representation for either a right or a
left spin-operator of the inter-dimer part of the exchange into (\ref{e2})
leads to single-hole exchange-scattering terms $\propto J_{1,2,3}$.  
Summing all contributions we arrive at the Hamiltonian
\begin{eqnarray}
H&=&-t_1 \sum_{n, \sigma} a_{1,n,\sigma}^\dagger
a_{2,n,\sigma}^{\phantom\dagger} + h.c. \nonumber \\
&+&\frac{t_2}{2} s^2 \sum_{n, \sigma} a_{2,n, \sigma}^\dagger a_{1, n-1,
\sigma}^{\phantom\dagger}+h.c. \nonumber \\
&+&\frac{t_3}{2} s^2 \sum_{j, n, \sigma} a_{j, n, \sigma}^\dagger a_{j,
n-1, \sigma}^{\phantom\dagger}+h.c. \nonumber \\
&+& t_2 s \sum_n {\bf t}_n^\dagger \left 
({\bf S}_{n+1, n}^{21}-{\bf S}_{n-1,
n}^{12}\right)+h.c. \nonumber \\
&+&\frac{J_2}{2} s \sum_n {\bf t}_n^\dagger \left ({\bf S}_{n-1,n-1}^{22} 
- {\bf S}_{n+1,n+1}^{11} \right)+h.c. \nonumber \\
&+&t_3 s \sum_{j, n}{\bf t}_n^\dagger (-1)^{j-1} \left({\bf S}_{n-1,
n}^{jj}+{\bf S}_{n+1,n}^{jj}\right)+h.c. \nonumber\\
&+& \frac{J_3}{2} s \sum_{j, n}{\bf t}_n^\dagger (-1)^{j-1} 
\left({\bf S}_{n-1,n-1}^{jj}+{\bf S}_{n+1,
n+1}^{jj}\right)+h.c. \nonumber\\
&-& \frac{t_2}{2} \sum_{n, \sigma} {\bf t}_{n-1}^\dagger {\bf
t}_{n}^{\phantom\dagger} a_{2,n,\sigma}^\dagger a_{1, n-1,
\sigma}^{\phantom\dagger} +h.c. \nonumber\\
&+&\frac{t_3}{2} \sum_{j, n, \sigma} {\bf t}_{n-1}^\dagger {\bf
t}_{n}^{\phantom\dagger} a_{j, n, \sigma}^\dagger a_{j, n-1,
\sigma}^{\phantom\dagger}+h.c. \nonumber \\
&+&t_2 \sum_{n} i \left ({\bf t}_n^\dagger \times {\bf
t}_{n+1}^{\phantom\dagger} \right ) {\bf S}_{n+1, n}^{21}+h.c. \nonumber \\
&-&\frac{J_2}{2}\sum_n i \left ({\bf t}_n^\dagger \times {\bf
t}_{n}^{\phantom\dagger} \right ) \left ({\bf S}_{n-1, n-1}^{22}
+{\bf S}_{n+1,
n+1}^{11}\right) \nonumber\\
&-&t_3 \sum_{j, n} i \left ({\bf t}_n^\dagger \times {\bf
t}_{n+1}^{\phantom\dagger} \right ) {\bf S}_{n+1, n}^{jj} 
+h.c. \nonumber \\
&-&\frac{J_3}{2} \sum_{j, n} i \left ({\bf t}_n^\dagger \times {\bf
t}_{n}^{\phantom\dagger} \right ) \left ({\bf S}_{n+1, n+1}^{jj}
+{\bf S}_{n-1,
n-1}^{jj}\right). \label{e16}
\end{eqnarray}
with
${\bf t}_n^\dagger=(t_{x,n}^\dagger,t_{y,n}^\dagger, t_{z,n}^\dagger)$.
This Hamiltonian is spin-rotationally invariant which is consistent 
with the ground state displaying no long-range magnetic order. 
According to the LHP approximation we have replaced the singlet 
operators by $C$-numbers, i.e. $s$. Within the LHP approach $s=1$ 
to lowest order. Regarding the single-hole Hamiltonian we improve 
upon this by assuming that the condensate density of the singlet is 
determined by satisfying the hardcore constraint (\ref{e5}) on the
average, i.e. by setting
$s_n^{\phantom\dagger}=s_n^\dagger =\langle s_n^{\phantom
\dagger}\rangle=s$ with
\be
s^2=1-\sum_\alpha \langle t_{\alpha, n}^{\dagger}
t_{\alpha, n}^{\phantom\dagger} \rangle =1-\frac{3}{N}\sum_q v_q^2.
\label{e14}
\ee
In principle this equation can be used to determine the LHP
magnon-dispersion selfconsistently which however will not be done here. For
the particular case of a ladder, i.e. $J_2=t_2=0$, Hamiltonian (\ref{e16})
agrees with refs. \cite{eder98,Sushkov99}. See ref. \cite{Vojta99} for a
related representation of the bilayer HT$_C$-cuprates.

\section{Single-hole Green's function}
\begin{figure}[tb]
\centerline{\psfig{file=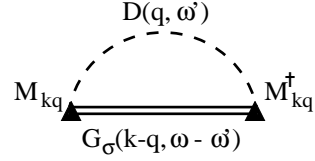,width=4cm,angle=-90}}
\vspace{0.3cm}
\caption{SCBA self-energy.}
\label{fig2}
\end{figure}

In this section we evaluate the retarded pseudo-fermion, and 
physical-fermion Green's function
\begin{eqnarray}\label{wb100}
G_\sigma(k,t) = && -i\Theta(t)
\langle D|\{\psi_{k,\sigma}^{\phantom\dagger}(t),
\psi_{k,\sigma}^\dagger\}|D\rangle
\\ \label{e25}
G_{\sigma}^c(k,t)= && -i\Theta(t)
\langle D|\{\phi_{k,\sigma}^{\phantom\dagger}(t),
\phi_{k,\sigma}^\dagger\}|D\rangle,
\end{eqnarray}
where $\psi_{k, \sigma}^\dagger=(a_{2,k,\sigma}^\dagger , a_{1,k,
\sigma}^\dagger)$ and $\phi_{k,\sigma}^\dagger=({\hat c}_{1,k,
\sigma}^\dagger,{\hat c}_{2,k,\sigma}^\dagger)$, i.e. both,
$G_\sigma(k,t)$ and $G_{\sigma }^c(k,t)$ are 2$\times$2 matrices. We 
proceed via standard diagrammatic techniques to to evaluate the Green's 
functions. To this end approximations have to be made. First, and in 
the limit of large dimerization we find\cite{cjup}, that renormalization 
effects due to the two-triplet vertices in (\ref{e16}) are of minor 
importance and we will discard them in the following. Fourier
transforming the corresponding simplified Hamiltonian and introducing 
the Bogoliubov representation of the
triplets, i.e. (\ref{e6}), we obtain $H=H_0+V$ with
\begin{eqnarray}
H_0&=&\sum_{k, \sigma} \psi_{k, \sigma}^\dagger E_k \psi_{k,
\sigma}^{\phantom\dagger}\nonumber \\
V&=&\frac{1}{\sqrt{N}}\sum_{k,q,\sigma_1, \sigma_2} {\bf
\gamma}_q^\dagger {\boldmath\tau}_{\sigma_1, \sigma_2}^{\phantom\dagger}
\psi_{k-q,\sigma_1}^\dagger M_{kq} \psi_{k, \sigma_2}, 
\label{e19}
\end{eqnarray}
where
\begin{equation}\label{e20}
E_k = \pmatrix{\mu_k &
\epsilon_k^{\phantom\star} \cr \epsilon_k^\star & \mu_k}\qquad
M_{kq}=\pmatrix{m_{1, kq}^{\phantom\star} &
m_{2, kq}^{\phantom\star} \cr -m_{2, kq}^\star & -m_{1, kq}^\star}
\end{equation}
with
\begin{eqnarray}\label{e20_1}
\epsilon_k= && -t_1+s^2 \frac{t_2}{2}e^{i k} 
\nonumber\\
\mu_k= && s^2 t_3 \cos k
\nonumber\\
m_{1, kq}= && u_q s
(\frac{J_2}{4}e^{i q}-t_3 \cos{(k-q)}-\frac{J_3}{2}\cos q )
\nonumber\\
&&+v_q s (\frac{J_2}{4}-t_3
\cos{k}-\frac{J_3}{2} \cos q )
\nonumber\\
m_{2, kq}=&&\frac{t_2}{2} s ( u_q e^{i (k-q)}-v_q e^{ik}).  
\label{e20_2}
\end{eqnarray}
The structure of (\ref{e19}-\ref{e20_1}) is reminiscent of the Hamiltonian
which occurs in the pseudo-particle description of the single-hole dynamics
in an antiferromagnetic (AFM) spin-background which has been studied
extensively in the context of the HT$_C$-compounds
\cite{Schmitt-Rink88,Kane89,Martinez91,Liu92}.  These studies have 
shown the so-called selfconsistent-Born-approximation (SCBA) to the 
pseudo-fermion Green's function to be in satisfactory agreement with 
information available from exact-diagonalization (ED) studies on finite 
lattices. The SCBA amounts to a non-crossing, infinite order resummation
of fermion-boson scattering-diagrams to the pseudo-fermion self-energy, 
neglecting vertex corrections as depicted in fig. \ref{fig2}.  In an AFM 
spin-background the neglect of vertex corrections can be justified 
in the limit of high coordination 
number\cite{Martinez91,Liu92,Brinkman70}.  Regarding a
dimer-state, i.e. $|D\rangle$, we are unaware of a similar simplification.
However, at the one-loop level we have checked numerically that the
inclusion of vertex corrections leads to relatively minor changes
only\cite{cjup}.

The SCBA self-energy $\Sigma_\sigma(k,\omega)$ of figure \ref{fig2} is
evaluated in terms of the bare pseudo-fermion Green's function
$G_\sigma^0(k,\omega)$ with respect to $H_0$ of (\ref{e19}), i.e.
\be
G_\sigma^0(k,\omega)=\frac{1}{(\omega-\mu_k)^2-|\epsilon_k|^2}
\pmatrix{\omega-\mu_k
& \epsilon_k^\star \cr \epsilon_k^{\phantom\star} &\omega-\mu_k}
\label{e21}
\ee
and the retarded triplet Green's-function $D_\alpha(k,t)= -i\Theta(t)
\langle D|[\gamma_{k,\alpha}^{\phantom\dagger}(t), 
\gamma_{k,\alpha}^\dagger
]|D\rangle$ the Fourier transform of which follows from Eq. (\ref{e7})
\be
D_\alpha(k,\omega)=\frac{1}{\omega-\omega_k}.
\label{e22}
\ee
where it is assumed that $\omega\equiv \omega+i \eta$ with $\eta
\rightarrow 0^+$ in (\ref{e21},\ref{e22}) as well as in all other 
retarded propagators in the remainder of this paper.  Inserting
(\ref{e21},\ref{e22}) into fig. \ref{fig2} we get\cite{crit_robert}
\begin{eqnarray}
\Sigma_\sigma(k, \omega) &= &\frac{3}{N}\sum_q M_{kq}^{\phantom\dagger}
 [G^0_\sigma(k-q,
\omega-\omega_q)^{-1}\nonumber \\&-&\Sigma_\sigma(k-q,w-w_q)]^{-1}
M_{kq}^\dagger
\label{e23}
\end{eqnarray}
from which the pseudo-fermion Green's function is obtained as usual via
$G_\sigma^{\phantom 0}(k,\omega)=[G_\sigma^0(k,\omega)^{-1}-
\Sigma_\sigma(k,\omega)]^{-1}$.

\begin{figure}[tb]
\centerline{\psfig{file=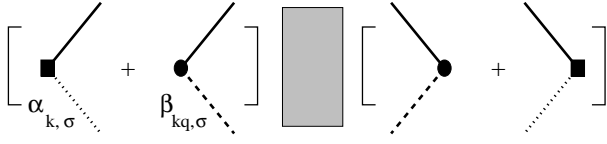,width=8cm,angle=-90}}
\vspace{0.5cm}
\caption{Structure of the physical Green's function 
$G_\sigma^c(k, \omega)$. Solid, dashed, and dotted lines refer to 
pseudo-fermion, triplet, and singlet Green's functions. Dashed box 
denotes reducible two particle vertex.
In LHP approximation dotted lines collapse to unity.}
\label{fig3a}
\end{figure}

The excitations of physical significance, eg. in a photoemission 
experiment, are the creation of holes via the two-particle operators
 $\phi_{k,\sigma}$\cite{Sushkov97}. The corresponding two-particle 
Green's function is depicted in fig. \ref{fig3a} where the vertices 
$\alpha_{k,\sigma}$ and $\beta_{kq,\sigma}$ correspond to the 
different ways by which the physical hole can be projected onto a 
pseudo-fermion and a dimer spin-state, i.e. a singlet or a triplet. 
This projection accounts for the ground-state quantum
fluctuations of the spin system.
\begin{eqnarray}
\alpha_{k,\sigma}= && 
s \langle D| \psi_{k, \overline{\sigma}}
\phi_{k,\sigma}| D \rangle =
\frac{\pm s}{\sqrt{2}} \pmatrix{1 & 0 \cr 0 & 1}
\label{e26}
\\
\beta_{kq,\sigma}= && \langle D|
(\gamma_q\psi_{k-q})_{\frac{1}{2}\frac{\mp 1}{2}}
\phi_{k,\sigma}|D \rangle =
\mp \sqrt{\frac{3}{2 N}} v_q \tau_z
\label{e27}
\end{eqnarray}
where, according to the LHP approximation, the singlet creator has been
replaced by a $C$-number and 
\begin{eqnarray}
\lefteqn{(\gamma_q\psi_{k-q})_{\frac{1}{2}\frac{\mp
1}{2}}=} &&\nonumber \\&&\frac{1}{\sqrt{3}}(\mp\gamma_{z,q}\psi_{k-q,\downarrow(\uparrow)} +
(\gamma_{x, q}\mp i \gamma_{y, q})\psi_{k-q, \uparrow(\downarrow)} )
\label{e28}
\end{eqnarray}
refers to the proper linear combination of spin-1/2 and spin-1 which 
ensures that the hole created by $\phi_{k,\sigma}$ has spin 
$S=\frac{1}{2}$ with $S_z=\pm \frac{1}{2}$.  In contrast to the 
vertices (\ref{e26},\ref{e27}) their respective products and squares 
are spin-independent.

\begin{figure}[tb]
\centerline{\psfig{file=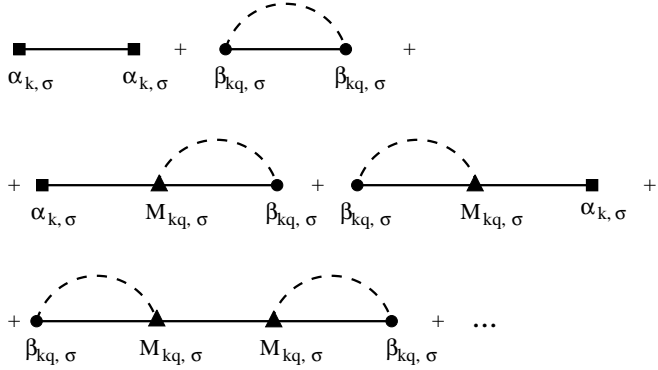,width=\linewidth,angle=-90}}
\vspace{0.1cm}
\caption{LHP+RPA approximation to the physical Green's function
$G_\sigma^c(k, \omega)$.}
\label{fig3}
\end{figure}

For a finite albeit small density of triplet-pairs in the 
ground state, i.e. $0<v_q\ll 1$, we proceed by a perturbative evaluation 
of the two-particle propagator of fig. \ref{fig3a} using the RPA as 
shown in fig. \ref{fig3}.  A comparable analysis of the effects of 
ground-state quantum fluctuations on the single-hole spectrum has been 
carried out in an AFM spin-background for the square-lattice 
$t$-$J$-model \cite{Sushkov97}.  Within the RPA the
physical Green's function is given by
\begin{eqnarray}
\lefteqn{G_\sigma^c(k,w) =} &&
\nonumber\\[5pt]
&&  \alpha_{k, \sigma} G_\sigma(k,w) \alpha_{k, \sigma} + \sum_q \beta_{kq,
\sigma} G_\sigma(k-q, \omega-\omega_q)\beta_{kq, \sigma}
\nonumber\\
&& +\pmatrix{
\alpha_{k, \sigma} G_\sigma(k,w) &  \sqrt{\frac{3}{N}}\sum_q \beta_{kq,
\sigma} G_\sigma(k-q,
\omega-\omega_q) M^\dagger_{kq}}
\nonumber \\[6pt]
&& \times \pmatrix{1 & -\frac{3}{N} M_{kq}  G_\sigma(k-q,\omega-\omega_q)
M^\dagger_{kq} \cr -G_\sigma(k,\omega)& 1}^{-1} 
\nonumber \\[6pt]
&& \times \pmatrix{\sqrt{\frac{3}{N}}\sum_q M_{kq} G_\sigma(k-q,
\omega-\omega_q) \beta_{kq, \sigma} \cr 
G_\sigma(k,\omega)\alpha_{k, \sigma}}
\label{e29}
\end{eqnarray}
where the element-by-element multiplication of the matrix and vector
entries in the matrix-product on the last three lines of this equation are
2$\times$2 matrices operations. To arrive at (\ref{e29}) the LHP dynamics
of the bond-bosons has been used which implies a constant factor of unity
only in case of the singlet propagator.

Concluding this section we note the sum rule
\be
\int_{-\infty}^\infty d\omega A_\sigma^c(k,
\omega)=\{\phi^{\phantom\dagger}_{k, \sigma}, \phi^{\dagger}_{k, \sigma}
\}=\frac{1}{2}
\pmatrix{1&0 \cr 0&1}
\label{e30}
\ee
for the exact physical Green's function with
$A_\sigma^c(k,\omega)=-\frac{1}{\pi}\mbox{Im}\; G_\sigma^c(k,\omega)$
at half-filling which differs from that for free fermions since the
$\hat{c}_{i,n, \sigma}^{(\dagger)}$ are Hubbard operators. Integrating
(\ref{e30}) we find that the  combined LHP and RPA approach is consistent
with this sum-rule yielding
\begin{eqnarray}
-\frac{1}{\pi}\int_{-\infty}^\infty \mbox{Im} G^c_\sigma(k,\omega)&&=
\alpha_{k,\sigma}^2+\sum_q \beta_{kq,\sigma}^2
\nonumber \\
&&=\frac{1}{2}\pmatrix{1&0 \cr 0&1} 
(s^2+\frac{3}{N}\sum_q v_q^2)
\label{eq31a}
\end{eqnarray} 
where the term in the last bracket is one due to (\ref{e14}).

\section{Dimerized chain spectra}
\label{sec6}
In the remaining sections of this work we will detail various results of 
the numerical solution of the SCBA and RPA equations. We begin with the
dimerized-chain limit of the $t_{1,2,3}$-$J_{1,2,3}$-model, i.e. with
$J_3=t_3=0$. In view of the non-zero dimerization it seems convenient to
visualize the lattice geometry as that of a linear chain with a unit cell
containing two electrons and a lattice constant $2a$, set equal to unity,
where $a$ is the inter-site distance. In contrast to this, existing ED
studies frequently employ a different representation in which each unit 
cell contains a single site only. This representation can be obtained by
introducing the fermion operators
\begin{eqnarray}
d_{\frac{k}{2}, \sigma}
=\frac{1}{\sqrt{2}}\left({\hat c}_{1, k,
\sigma}+e^{-i\frac{k}{2}}{\hat c}_{2,k,\sigma}\right),
\label{e31}
\end{eqnarray}
which we term $d$-electrons hereafter.
The $d$-Green's-function $G_\sigma^d(k, \omega)$ reads
\begin{eqnarray}
G_\sigma^d(k, \omega)&=&\frac{1}{2} [(G_\sigma^c(k,
\omega))_{11}
+(G_\sigma^c(k,\omega))_{22} \nonumber \\&+&e^{-i\frac{k}{2}}(G_\sigma^c(k,
\omega))_{12}+e^{i\frac{k}{2}}(G_\sigma^c(k,
\omega))_{21}], 
\label{e32}
\end{eqnarray}
with $A^d_\sigma(k, \omega)=-\frac{1}{\pi}\mbox{Im}\; G_\sigma^d(k,\omega)$
beeing the corresponding spectral function.

\begin{figure}[tb]
\centerline{\psfig{file=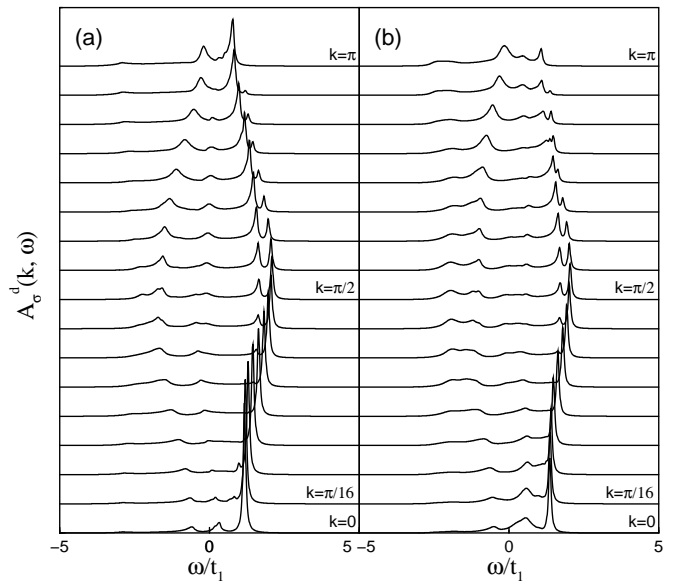,width=\linewidth,angle=-90}}
\caption{
Spectral function of the dimerized chain for various values of 
momentum $k=0$ to $\pi$ and $t_1=t_2$, $N=256$, and $\eta=0.05$. 
(a)  $J_0/t_1=1$, $\Delta/t_1=0.2$ and (b) $J_0/t_1=0.5$, 
$\Delta/t_1=0.1$.}
\label{fig6}
\end{figure}

Selfconsistent solutions for the SCBA can be obtained quite easily by
numerical iteration on fairly large lattices and for small, but 
finite imaginary broadening $\eta\ll 1$.
Convergence of the iteration is achieved within a few, typically 3-20, 
cycles depending on the size of the spin gap.  Moreover, finite size 
scaling analysis can be used to determine an approximate system size at 
which further increase in the number of lattice sites leads to no 
additional change in the pseudo-fermion Green's function at fixed 
$\eta$, i.e. the thermodynamic limit. Inserting SCBA Green's-functions 
from this limit into the RPA (\ref{e29}) we obtain typical spectra as 
displayed in fig. \ref{fig6}.  The number of sites is $512=2\times N$ 
where $N$ is the number of dimers. The dimerization of the hopping 
integrals and the exchange couplings have been chosen independently with 
$t_1=t_2$ and $\delta$ such as to result in a spin gap of 
$\Delta/t_1=0.2J_0/t_1$. Since $J_{^1_2}=J_0(1\pm\delta)$ one
finds $\Delta/t_1=[2 (\delta^2+\delta)]^{1/2}J_0/t_1$. Figure \ref{fig6}a)
(b)) refers to $J_0/t_1=1 (0.5)$ and $\Delta/t_1=0.2(0.1)$.  All spectra 
are particle-hole inverted, i.e. $\omega \rightarrow -\omega$, such as 
to place the first electron-removal state at the lowest binding-energy.

\begin{figure}[tb]
\centerline{\psfig{file=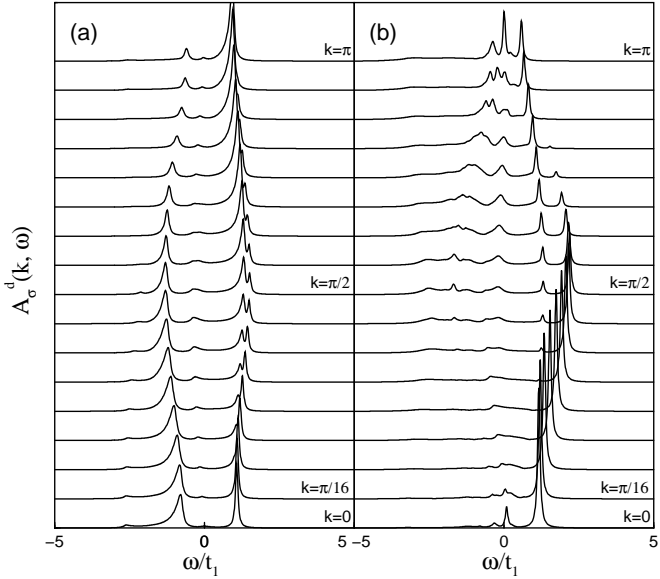,width=\linewidth,angle=-90}}
\caption{Spectral function of the dimerized chain for 
various values of momentum $k=0$ to $\pi$ and $t_1=t_2=J_0$,  
$N=256$, and $\eta=0.05$. (a) $\Delta/t_1=0.05$ and (b) $\Delta/t_1=0.7$.}
\label{fig7}
\end{figure}

The figure displays two dominant dispersive regions which are related 
to the first three terms in Hamiltonian (\ref{e16}). These terms allow 
for coherent hole-motion {\em without} triplet excitations of the 
spin-background and lead
to two tight-binding-type bands which are accounted for by $H_0$ in
Hamiltonian (\ref{e19}). These tight-binding states are renormalized and
broadened upon coupling to multi triplet-excitations via the SCBA. In
particular the states of high binding-energy are broadened quite strongly.
Comparing fig. \ref{fig6}a) vs. b) it is obvious that the amount of the
spectral redistribution depends on the relative size of $J_0/t_1$, where
decreasing $J_0/t_1$ implies a larger incoherent part of the spectrum.
Additionally the dispersion flattens as $J_0/t_1$ decreases.

Finite-size analysis of the two sharp structures at the high-energy edge of
the spectrum in figure \ref{fig6}a) (b)) reveals that for any finite
spin-gap these structures refer to poles of the Green's function on the
real axis which remain separated by a gap from the continuum of 
multi-magnon shake-offs. The spectral weight of these poles remains 
finite in the thermodynamic limit. Therefore, the first electron 
removal state is of a quasi-particle type. The first of these two 
poles, i.e. at higher binding-energy, can be traced back to a 
quasi-particle excitation in the pseudo-fermion SCBA-spectrum which is 
shadowed in the physical fermion Green's function. The second pole at 
lowest binding energy, i.e. the first electron removal state, arises from 
a zero of the determinant of the RPA-matrix propagator in (\ref{e29}). 
This can be interpreted in terms of a bound state between the 
aforementioned pseudo-fermion quasi-particle and the ground-state quantum 
fluctuations of the spin-system.

Figure \ref{fig7} refers to the dependence of the spectra on the spin-gap
displaying two additional values of  $\delta$ different from that used
in fig. \ref{fig6}. We find that the gap between the first electron-removal
state and the second pole increases upon increasing the spin gap. Moreover
the dispersion decreases for smaller values of $\delta$. For
$\Delta/t_1=0.05J_0/t_1$ and $J_0=t_1$, fig. \ref{fig6}a), an intense 
low energy band arises. The dispersion of the two dominant spectral
regions at $\delta\ll 1$ are reminiscent of similar findings for the
infinite-$U$ Hubbard-chain at half filling \cite{Sorella92}.

\begin{figure}[bt]
\centerline{\psfig{file=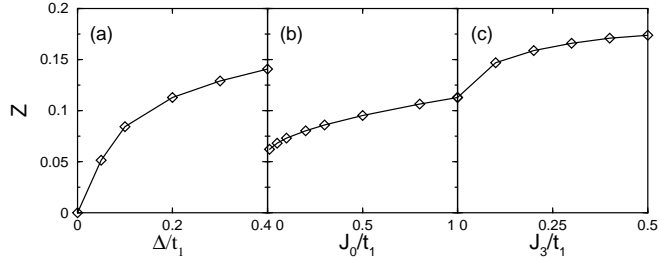,width=\linewidth,angle=-90}}
\caption{Quasiparticle weight $Z(k=\pi/2)$ as a function (a) 
of $\Delta/t_1$ for  $J_0=t_1$, $J_3=0$,
 (b) of $J_0/t_1$ for $\Delta/t_1=0.2J_0/t_1$, i.e. $\delta=const.$, 
 $J_3=0$, and (c) of $J_3/t_1$ for $\delta$ as in (b), $J_0=t_1$.  
 $N=256$, $t_1=t_2$.}  
\label{fig8}
\end{figure}

Figure \ref{fig8} shows the weight, i.e. the $Z$-factor, of the
quasi-particle pole at wave vector $k=\pi/2$ as a function of various
parameters, $\delta$, $J_0/t_1$, and $J_3/t_1$. The $Z$-factor has been
determined by fitting a Lorentzian to the quasi-particle peak.  The
renormalization effects are quite strong for those parameters we have
considered as $Z$ is reduced substantially from the non interacting 
value of one half (\ref{e30}).  This figure demonstrates that a finite 
spin-gap stabilizes the quasi-particle excitations, while approaching the 
Heisenberg point, i.e. $\delta=0$, $J_3=0$, as in fig. \ref{fig8}a) 
suppresses the $Z$-factor due to the decrease in energy of available 
triplet excitations. In fig. \ref{fig8}b) the dimerization is kept at a 
constant value to result in $\Delta/t_1=0.2J_0/t_1$ with only $J_0/t_1$ 
varying. This leads to almost no change in the $Z$-factor.  Increasing 
frustration, as in fig. \ref{fig8}c) increases both, the average absolute 
value of the vertex-function $M_{kq}$ of (\ref{e20}) and the spin-gap 
$\Delta$. These effects compete leading to only a slight increase of the 
$Z$-factor as a function of $J_3$. Here $\delta$ is still kept at the 
constant value of fig. \ref{fig8}b).  While the single-hole case does not 
resemble a finite hole concentration it is tempting to relate the finite 
$Z$-factor in the presence of a spin-gap to a Luther-Emery-liquid 
\cite{Luther74} scenario in contrast to the Luttinger-liquid 
\cite{Haldane80} at vanishing spin-gap.

\begin{figure}[bt]
\centerline{\psfig{file=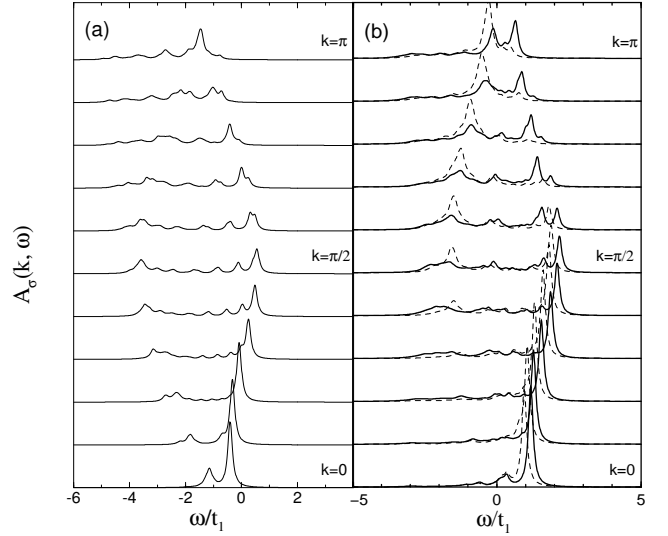,width=\linewidth}}
\caption{Comparison of spectra for a dimerized $20$-site chain for various
values of momentum $k=0$ to $\pi$ and $t_1=t_2=J_0$, $\delta=0.048$. (a) ED
data, reproduced from ref.$^{31}$
(b) solid line SCBA+RPA spectrum $A^d_\sigma(k, \omega)$, dashed line
SCBA-only spectrum $A^a_\sigma(k, \omega)$. $\eta=0.05$. (a) and (b) are
shifted relative to each other due to differing zeros of energy.}
\label{fig5}
\end{figure}

Next we turn to a comparison of our diagrammatic approach with results
obtained from exact diagonalization (ED) of finite chains. To this end, we
contrast a single-hole spectrum for $\delta=0.048$ and $t_1=t_2=J_0$
reproduced from the work of Augier and collaborators \cite{augier97} 
in fig.
\ref{fig5}a) against a physical fermion Green's function obtained via
(\ref{e29}) using an identical number of sites in fig. \ref{fig5}b).
The agreement, albeit  qualitative, is rather satisfying. 
In \ref{fig5}b) the
thin dashed line refers to the pseudo-fermion spectrum. Obviously the
similarity of this function to the ED result is less convincing thereby
demonstrating the relevance of ground-state quantum fluctuations. The
difference between the pseudo- and the physical-fermion spectrum is
particularly evident in the incoherent part at wave vectors larger than
$\pi/2$.

\begin{figure}[tb]
\centerline{\psfig{file=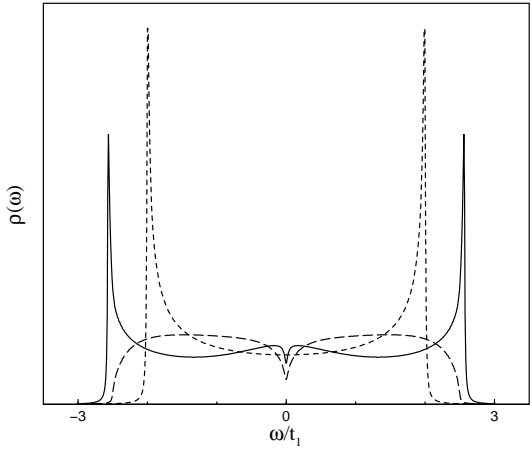,width=7cm,angle=-90}}
\caption{$t$-model DOS for $N=256$ and $t_1=t_2$. Solid, long dashed, and
short dashed line refer to the SCBA+RPA, i.e.  $\rho_d(\omega)$, SCBA-only,
i.e.  $\rho_a(\omega)$, and Brinkman-Rice result$^{57}$.}
\label{fig4}
\end{figure}

Closing this section we comment on the so-called dimerized $t$-model limit,
i.e. $t_1=t_2$, $J_1\gg J_2$ with $J_1\rightarrow 0$, and $t_3=J_3=0$. In
this case the ground state of the spin system is a perfect dimer
product-state while the only energy scale of the system is the hopping
amplitude $t=t_1$.  The local density of states $\rho(\omega)$ for a single
hole in one dimension is known to be $\rho(\omega)=1/\sqrt{\omega^2-4 t^2}$
which is independent of the particular spin-background \cite{Brinkman70}.
Calculating $\rho_{d(a)}(\omega)=\frac{1}{2 N}\sum_k A^{d(a)}_\sigma(k,
\omega)$ using the representation (\ref{e31}) and (\ref{e32})
we obtain the result shown in fig \ref{fig4}. The SCBA+RPA approach shows
some deviation from the exact result regarding the low-frequency regime and
the overall band-width.  The latter effect is reminiscent of similar
findings from analysis of the 2D AFM $t$-model\cite{Kane89}.  The pure SCBA
pseudo-fermion spectrum shows little resemblance with the exact result.

\section{Spin-ladder spectra}
\label{sec7}
 In this section we turn to the case of $J_2=t_2=0$ which describes a
two-leg spin ladder. In contrast to the definition (\ref{e31}) of the
fermions to account for the 1D representation, the unit cell remains
identical to that of the lattice topology of fig. \ref{fig1}. The
pseudo-fermion operators on a rung can be labeled according to their
(anti)bonding symmetry and the reciprocal space is quasi two-dimensional
with a wave vector ${\bf k}=(k,k_y)$ where $k_y=(\pi)0$ which refers to the
(anti)bonding state
\be
b_{k, k_y, \sigma}^\dagger=\frac{1}{\sqrt{2}} \left( a_{2, k,
\sigma}^\dagger + e^{i k_y} a_{1, k, \sigma}^\dagger \right).
\label{e34}
\ee
Expressing $G_\sigma^c(k,\omega)$ in terms of $b_{k, k_y, \sigma}$ renders
the Green's function diagonal because the Hamiltonian conserves parity and
the operators (\ref{e34}) exhibit a different signature with respect to
reflections at a plane perpendicular to the rungs
\cite{gopalan94}. The pseudo and physical
fermion spectral functions are labeled accordingly, i.e.
$A^{a}_{0(\pi),\sigma}(k,\omega)$ and $A^{c}_{0(\pi),\sigma}(k,\omega)$,
where the subscript $0(\pi)$ refers to the (anti)bonding symmetry.

Conservation of parity leads to an important constraint
regarding the pseudo-fermion self-energy. Since the spin-triplet has
odd parity, scattering of a pseudo-fermion off a triplet can occur only
via a transition between the bonding and the antibonding fermionic states
which have different parity. This implies a certain 'robustness' of the 
bare
pseudo-fermion bands of $H_0$ of (\ref{e19}) against renormalization by
$V$ because of the finite energy gap $2t_1$ between the bare bands given by
$\epsilon_{\pi(0)} (k)= (-)t_1+t_3 \cos (k)$. Therefore, perturbative
analysis of the single-hole dynamics on a ladder using methods
complementary\cite{Sushkov99,endres96} as well as
related\cite{eder98,crit_robert} to that of this work have led to rather
good agreement with numerical studies of finite system.

\begin{figure}[bt]
\centerline{\psfig{file=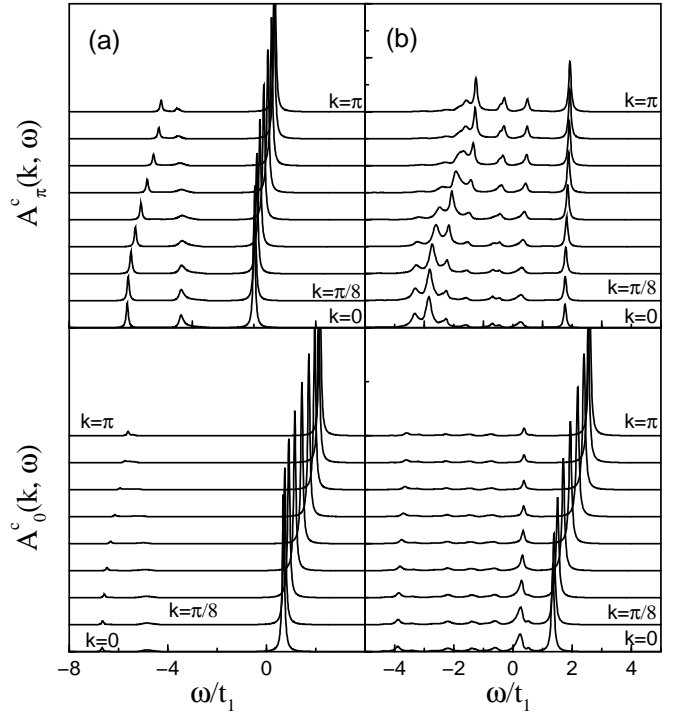,width=\linewidth,angle=-90}}
\caption{Spectral function $A^c_{0(\pi),\sigma}$ of the bonding ($0$) and
antibonding $(\pi)$ bands of the two-leg ladder for various values of
momentum $k=0$ to $\pi$. $N=256$, $\eta=0.1$. (a) $t_1=t_3$, $J_1/t_1=5$,
$J_3/t_1=0.5$. (b) $t_1=t_3$, $J_1/t_1=1$, $J_3/t_1=0.1$.}
\label{fig10}
\end{figure}

In fig. \ref{fig10}a) the (anti)bonding spectra are shown for a case of
strong intra-rung exchange $J_1/t_1=5$, $J_3/t_1=0.5$. The number of rungs
is set to $N=256$ for which finite-size effects are negligible. The
ground-state in this case is close to a pure singlet product-state with
almost no quantum fluctuations. The spin-gap is relatively large,
i.e. $\Delta/t_1\approx 4.5$. The spectrum clearly displays the remnants of
the (anti)bonding bare bands and the intensity of the incoherent
contributions to the spectrum are weak.  Next, fig. \ref{fig10}b), with
$J_1/t_1=1$, $J_3/t_1=0.1$, refers a smaller gap which accounts for the
extra incoherence of the spectrum and the less correspondence between the
bare and the renormalized bands. Moreover the high energy parts of the two
bands intersect.

Unfortunately, at the isotropic point, i.e. $J_1=J_3$, the LHP
approximation breaks down, with the spin-gap closing for $J_3\geq J_1
/2$. Yet, from series expansion
\cite{Reigrotzki94,Barnes93,Oitmaa96,Pieka98} and DMRG \cite{White94}
it is known, that $\Delta/J\approx 0.5$ in that case. To account for
this deficiency of the LHP we keep the isotropic set of parameters
in all of (\ref{e19}-\ref{e20}) however we adjust $J_3$ in
(\ref{e7}-\ref{e10}) such as to result in $\Delta/J_1 =0.5$.
We believe that this modification provides an approximate description
of the renormalization of $\omega_k$ which occurs beyond the LHP
approach.

\begin{figure}[bt]
\centerline{\psfig{file=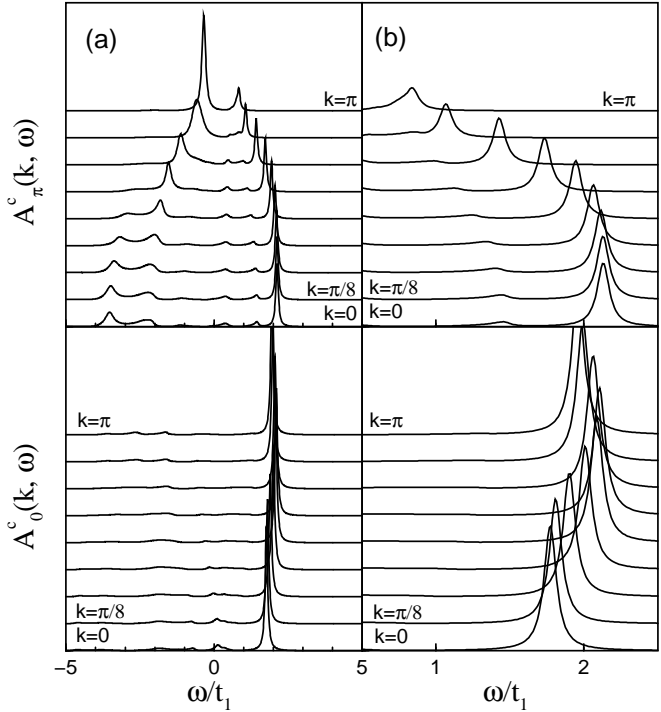,width=\linewidth,angle=-90}}
\caption{Spectral function $A^c_{0(\pi),\sigma}$ of the bonding ($0$) and
antibonding $(\pi)$ bands of the two-leg ladder for various values of
momentum $k=0$ to $\pi$ at the isotropic point  $t_1=t_3=J_1=J_3$. $N=256$,
$\eta=0.1$. (a) complete spectrum (b) details at low binding-energy.}
\label{fig10a}
\end{figure}

Fig. \ref{fig10a}a) and b) display the case of isotropic exchange and
hopping, i.e. $t_1=t_3=J_1=J_3$. Here the renormalization of the free
particle bands is very strong. The bonding band at highest frequency, is
very flat. In the antibonding band the intensity at highest energy is a
composite excitation of states from the bonding band and a single triplet.
Its relatively large intensity is a result of parity conservation, which
inhibits intraband scattering. In addition a lower energy feature,
reminiscent of the free antibonding band, can be observed at in the 
vicinity
of $k=\pi$.  Fig. \ref{fig10a}b) shows a stretched frequency window with 
the
two bands of lowest binding energy. Interestingly, the maximum of the
dispersion is found {\em off} the momenta $k=0$ or $\pi$. This is 
consistent
with ED studies \cite{Troyer96} and series expansions
\cite{oitmaa99}.

\begin{figure}[bt]
\centerline{\psfig{file=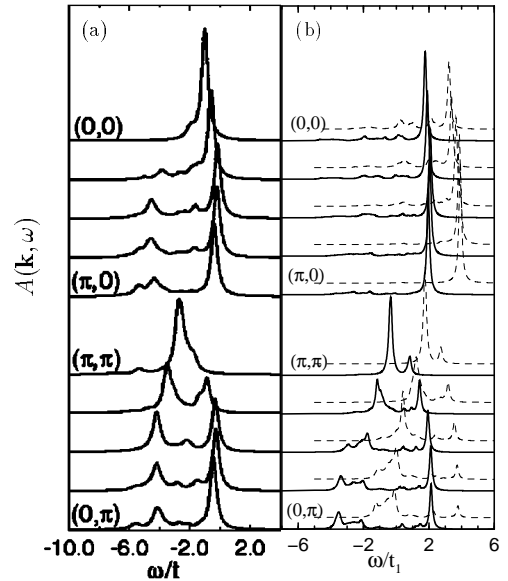,width=0.75\linewidth}}
\caption{Comparison of spectra for an $N=8$-rung two-leg spin-ladder for
various values of momentum $k=0$ to $\pi$. $t_1=t_3=J_1=J_3$, and
$\eta=0.1$.  (a) ED data, reproduced from ref.$^{30}$
(b) solid line SCBA+RPA spectrum $A^c_{\sigma}({\bf k},\omega)$, 
dashed line
SCBA-only spectrum $A^a_{\sigma}({\bf k},\omega)$.  For better visibility
the SCBA-only spectra have been shifted by $\Delta\omega/t=2$ along 
the $x$, an arbitrary amount along the $y$ axis.}
\label{fig9}
\end{figure}

Analogous to the case of the linear chain we contrast our diagrammatic
calculation against ED analysis. In fig. \ref{fig9}a) a spectrum reproduced
from the work of Haas and collaborators \cite{Haas96} is compared to a
physical-fermion spectrum obtained from the SCBA+RPA approach. The latter
spectrum is shown in fig. \ref{fig9}b). It compares well with the ED 
result.
Finally the dashed line in \ref{fig9}b) displays the SCBA, i.e.
pseudo-fermion spectrum. The latter spectrum lacks the large intensity in
the high energy region of the antibonding band and the displacement of the
maximum of the dispersion at low binding energy mentioned in the previous
paragraph.  This demonstrates that an inclusion of ground-state
quantum-fluctuations is mandatory for a proper description of the
single-hole dynamics in this regime of parameters.

\section{Conclusion}
In conclusion we have investigated the single-hole excitations at half
filling in dimerized and frustrated spin-chains and ladders in the
spin-dimer phase. Based on an analytic approach we have provided a
physically intuitive picture of the single-hole dynamics as resulting 
from a combination of the dimerization of the spin background and the 
scattering of the holes by massive triplet-modes of the spin system. 
Due to the spin-gap we find, that the single-hole excitations are 
quasi-particle like though strongly renormalized.  The impact of ground 
state quantum fluctuations of the dimer state on the hole-spectrum has 
been elucidated. Our results compares well with numerical diagonalization 
of finite systems, both for chains and ladders.  We hope that our work 
will promote future experimental studies to obtain angular resolved 
photoemission spectra on the novel chain-and ladder-type transition-metal 
oxides in an energy window comparable to the magnetic exchange coupling.

It is a pleasure to acknowledge helpful discussions with G. Baskaran, M.
Brunner, R. Eder, and O. Sushkov. This research was supported in part 
by the
Deutsche Forschungsgemeinschaft under Grant No.  BR 1084/1-1.


\begin{references}
\bibitem{crit_robert}
In \cite{eder98} a version of eqn. (\ref{e23}) has been considered
for the spin ladder with the term '$-w_q$' in the argument of the
selfenergy lacking. We are unable to reproduce this form of the SCBA.
In turn, the SCBA-spectra found in \cite{eder98} differ considerably
from ours.
\end{references}
\end{document}